\documentclass[12pt]{iopart}
\usepackage{epsf}
\usepackage{graphicx}
\begin{document}
\def\ra{\rangle}
\def\la{\langle}
\def\be{\begin{eqnarray}}
\def\ee{\end{eqnarray}}
\def\ha{{\hat a}}
\def\hb{{\hat b}}
\def\hu{{\hat u}}
\def\hv{{\hat v}}
\def\hc{{\hat c}}
\def\hd{{\hat d}}
\def\no{\noindent}\def\non{\nonumber}
\def\hi{\hangindent=45pt}
\def\v{\vskip 12pt}

\newcommand{\bra}[1]{\left\langle #1 \right\vert}
\newcommand{\ket}[1]{\left\vert #1 \right\rangle}

\review[Physics of Information] {The Physics of Information}
\author{Christoph Adami}
\address{Quantum Computing Technologies
Group, Jet Propulsion Laboratory MS 126-347, \\ California Institute of Technology
Pasadena, CA~91109}

\begin{abstract}
Information theory is a statistical theory concerned with the relative state of
detectors and physical systems. As a consequence, the classical framework of
Shannon needs to be extended to deal with quantum detectors, possibly moving at relativistic
speeds, conceivably within curved space-time. Considerable progress toward such a theory has
been achieved in the last ten years, while much is still not understood. This
review recapitulates some milestones along this road, and speculates about future
ones.
\end{abstract}

\section{Entropy and Information: Classical Theory}
Since Shannon's historical pair of papers~\cite{Shannon48}, information theory has
changed from an engineering discipline to a full-fledged theory within
physics~\cite{Landauer91}.  While a considerable part of Shannon's theory deals
with communication channels and codes~\cite{CoverThomas}, the concepts of entropy
and information he introduced are crucial to our understanding of the physics of
measurement, and turn out to be more general than thermodynamical entropy. Thus,
information theory represents an important part of statistical physics.

When discussing the relationship between information theory and statistical
physics, it is impossible not to mention Jaynes' work on the
subject~\cite{Jaynes1957}, who realized that optimal inference (that is, making
predictions based on available information) involves choosing probability
distributions that maximize Shannon's entropy. In this manner, he was
able to {\em justify} certain parts of statistical physics using probability
theory. The general point of view promulgated here goes beyond that. It is argued
that information theory is a unifying framework which can be used to describe those
circumstances in nonequilibrium physics that involve an observer and an observed.

In the following, I present an overview of some crucial aspects of entropy and
information in classical and quantum physics, with extensions to the special and
general theory of relativity. While not exhaustive, the treatment is at an
introductory level, with pointers to the technical literature where appropriate.

\subsection{Entropy}
The concepts of entropy and information quantify the ability of observers to make
{\it predictions}, in particular how well an observer equipped with a specific
measurement apparatus can make predictions about another physical system. Shannon
entropy (also known as {\em uncertainty}) is defined for mathematical objects
called {\em random variables}. A random variable $X$ is a system that can take on a
finite number of discrete states $x_i$, where $i=1,...,N$ with probabilities $p_i$.
Now, physical systems are not mathematical objects, nor are their states necessarily
discrete. However, if I want to quantify my uncertainty about the state of a
physical system, then in reality I need to quantify my uncertainty about the {\em
possible outcomes of a measurement of that system}. In other words, my maximal
uncertainty about a system is not a property of the system, but rather a property
of the measurement device with which I am about to examine the system. If my
measurement device, for example, is simply a ``presence-detector'', then the
maximal uncertainty I have about the physical system under consideration is 1 bit,
which is the amount of {\em potential information} I can obtain about that system.

Thus, the entropy of a physical system is undefined if we do not specify the device
that we are going to use to reduce that entropy. A standard example for a random
variable that is also a physical system is the six-sided even die. Usually, the
maximal entropy attributed to this system is $\log_2(6)$ bits. Is this all there is
to know about it? If it is a physical system, the die is made of molecules and
these can be in different states depending on the temperature of the system. Are
those knowable? What about the state of the atoms making up the molecules? They
could conceivably provide labels such that the number of states is in reality much
larger. What about the state of the nuclei? Or the quarks and gluons inside those?

This type of thinking makes it clear that indeed we cannot speak about the entropy
of an isolated system without reference to the coarse-graining of states that is
implied by the choice of detector. And even though detectors exist that record
continuous variables (such as, say, a mercury thermometer), each detector has a
finite resolution such that it is indeed appropriate to consider only the {\em
discrete} version of the Shannon entropy, which is given in terms of the
probabilities $p_i$ as\footnote{From now on, I shall not indicate the basis of the
logarithm, which only serves to set the units of entropy and information (base 2,
e.g., sets the unit to a ``bit'').}

\be H(X) = - \sum_i^N p_i \log p_i\;.  \label{entropy}\ee

For any physical system, how are those probabilities obtained? In principle, this
can be done both by experiment and by theory. Once I have defined the $N$ possible
states of my system by choosing a detector for it, the {\it a priori} maximal
entropy is defined as
\be 
H_{\rm max}= \log N\;. 
\ee
Experiments using my detector can now sharpen my knowledge of the system.  By
tabulating the frequency with which each of the $N$ states appears, we can estimate
the probabilities $p_i$. Note, however, that this is a biased estimate that
approaches the true entropy Eq.(\ref{entropy}) only in the limit of an infinite
number of trials. On the other hand, some of the possible states of the system (or
more precisely, possible states of my detector interacting with the system) can be
eliminated by using some knowledge of the physics of the system. For example, we
may have some initial data about the system. This becomes clear in particular if
the degrees of freedom that we choose to characterize the system with are position,
momentum, and energy, i.e., if we consider the {\em thermodynamical entropy} of the
system (see below).

\subsection{Conditional Entropy}
Let us look at the basic process that reduces uncertainty: a measurement.  When
measuring the state of system $X$, I need to bring it into contact with a system
$Y$. If $Y$ is my measurement device, then usually I can consider it to be
completely known (at least, it is completely known with respect to the degrees of
freedom I {\em care} about).  In other words, my device is in a particular state
$y_0$ with certainty. After interacting with $X$, this is not the case anymore. Let
us imagine an interaction between the systems $X$ and $Y$ that is such that \be x_i
y_0 \rightarrow x_iy_i\ \ \ \ \ i=1,...,N\;, \label{cmeas} \ee that is, the states
of the measurement device $y_i$ end up reflecting the states of $X$.  This is a
perfect measurement, since no state of $X$ remains unresolved.  More generally, let
$X$ have $N$ states while $Y$ has $M$ states, and let us suppose that $M<N$.  Then
we can imagine that each state of $Y$ reflects an {\em average} of a number of
$X$'s states, so that the probability to find $Y$ in state $y_j$ is given by $q_j$,
where $q_j=\sum_i p_{ij}$, and $p_{ij}$ is the joint probability to find $X$ in
state $x_i$ and $Y$ in state $y_j$. The measurement process then proceeds as \be
x_iy_0\rightarrow \langle x\rangle_j y_j \ee where \be \la x\ra_j=\sum_i p_{i|j}
x_i\;. \label{cond} \ee In Eq.(\ref{cond}) above, I introduced the {\em conditional
probability} \be p_{i|j} = \frac{p_{ij}}{q_j} \ee that $X$ is in state $i$ {\em
given} that $Y$ is in state $j$. In the perfect measurement above, this probability
was 1 if $i=j$ and 0 otherwise (i.e., $p_{i|j}=\delta_{ij}$), but in the imperfect
measurement, $X$ is distributed across some of its states $i$ with a probability
distribution $p_{i|j}$, for each $j$.

We can then calculate the {\em conditional entropy} (or remaining entropy) of the
system $X$ given we found $Y$ in a particular state $y_j$ after the measurement:
\be 
H(X|Y=y_j)= -\sum_i^N p_{i|j}\log p_{i|j}\;. 
\ee
This remaining entropy is guaranteed to be smaller than or equal to the
unconditional entropy $H(X)$, because the worst case scenario is that $Y$ doesn't
resolve {\em any} states of $X$, in which case $p_{i|j}$ = $p_i$. But since we
didn't know anything about $X$ to begin with, $p_i=1/N$, and thus $H(X|Y=y_j)\leq
\log N$.

Let us imagine that we did learn something from the measurement of $Y$, and let us
imagine furthermore that this knowledge is permanent. Then we can express our
new-found knowledge about $X$ by saying that we know the probability distribution
of $X$, $p_i$, and this distribution is {\em not} the uniform distribution
$p_i=1/N$. Of course, in principle we should say that this is a conditional
probability $p_{i|j}$, but if the knowledge we have obtained is permanent, there is
no need to constantly remind ourselves that the probability distribution is
conditional on our knowledge of certain other variables connected with $X$. We
simply say that $X$ is distributed according to $p_i$, and the entropy of $X$ is
\be 
H_{\rm actual}(X)=-\sum_i \log p_i \log p_i \;. \label{shannon} 
\ee
According to this strict view, all Shannon entropies of the form (\ref{shannon})
are conditional if they are not maximal. And we can quantify our knowledge about
$X$ simply by subtracting this uncertainty from the maximal one: \be I= H_{\rm
max}(X) - H_{\rm actual}(X)\;.\label{entdiff} \ee This knowledge, of course, is
{\em information}.

\subsubsection{Example: Thermodynamics}
We can view thermodynamics as a particular case of Shannon theory. First, if we
agree that the degrees of freedom of interest are position and momentum, then the
maximal entropy of any system is defined by its volume in phase space: \be H_{\rm
max} = \log \Delta \Gamma\;, \label{thermomaxent} \ee where $\Delta \Gamma =
\frac{\Delta p\Delta q}{k}$ is the number of states within the phase space volume
$\Delta p\Delta q$. Now the normalization factor $k$ introduced in
(\ref{thermomaxent}) clearly serves again to coarse-grain the number of states, and
should be related to the resolution of our measurement device. In quantum
mechanics, of course, this factor is given by the amount of phase space volume
occupied by each quantum state, $k=(2\pi\hbar)^n$ where $n$ is the number of
degrees of freedom of the system. Does this mean that in this case it is not my
type of detector that sets the maximum entropy of the system? Actually, this is
still true, only that here we assume a quantum mechanical perfect detector, while
still averaging over certain internal states of the system inaccessible to this
detector.

Suppose I am contemplating a system whose maximum entropy I have determined to be
Eq.~(\ref{thermomaxent}), but I have some additional information. For example, I
know that this system has been undisturbed for a long time, and I know its total
energy $E$, and perhaps even the temperature $T$. Of course, this kind of knowledge
can be obtained in a number of different ways. It could be obtained by experiment,
or it could be obtained by inference, or theory. How does this knowledge reduce my
uncertainty? In this case, we use our {\em knowledge of physics} to predict that
the probabilities $\rho(p,q)$ going into our entropy 
\be 
H(p,q)=-\sum_{\Delta
p,\Delta q} \rho(p,q)\log \rho(p,q) \label{therment}
\ee 
are given by the canonical
distribution\footnote{We set Boltzmann's constant equal to 1 throughout. This
constant, of course, sets the scale of thermodynamical entropy, and would end up
multiplying the Shannon entropy just like any particular choice of base for the
logarithm would.} \be \rho(p,q)= \frac1Z e^{-E(p,q)/T}\;, \ee where $Z$ is the
usual normalization constant, and the sum in (\ref{therment}) goes over all momenta
in the phase space volume $\Delta p\Delta q$. The amount of knowledge we have about
the system is then just the difference between these two uncertainties: \be I =
\log \Delta \Gamma - \log Z -\frac ET\;. \ee
\subsection{Information}
In Eq.~(\ref{entdiff}), we quantified our knowledge about the states of $X$ by the
difference between the maximal and the actual entropy of the system. This was a
special case because we assumed that after the measurement, $Y$ was in state $y_j$
with certainty, i.e, everything was known about it. In general, we can imagine that
$Y$ instead is in state $y_j$ with probability $q_j$ (in other words, we have some
information about $Y$ but we don't know everything, just as for $X$). We can then
define the {\em average conditional entropy} of $X$ simply as \be H(X|Y) = \sum_j
q_j H(X|Y=y_j) \label{condent} \ee and the information that $Y$ has about $X$ is
then the difference between the unconditional entropy $H(X)$ and
Eq.~(\ref{condent}) above, \be H(X:Y)=H(X)-H(X|Y)\;. \label{info} \ee The colon
between $X$ and $Y$ in the expression for the information $H(X:Y)$ is conventional,
and indicates that it stands for an entropy {\em shared} between $X$ and $Y$.
According to the strict definition given above, $H(X)=\log N$, but in the standard
literature $H(X)$ refers to the actual uncertainty of $X$ given whatever knowledge
allowed me to obtain the probability distribution $p_i$, i.e., Eq.~(\ref{shannon}).

Eq.~(\ref{info}) can be rewritten to display the symmetry between the observing
system and the observed: \be H(X:Y)=H(X)+H(Y)-H(XY)\;, \label{infosymm} \ee where
$H(XY)$ is just the joint entropy of both $X$ and $Y$ combined. This joint entropy
would equal the sum of each of $X$'s and $Y$'s entropy only in the case that there
are {\em no correlations} between $X$'s and $Y$'s states. If that would be the
case, we could not make any predictions about $X$ just from knowing something about
$Y$. The information (\ref{infosymm}), therefore, would vanish.

\subsubsection{Measurement Example} \label{measexample}
An instructive example illustrating the effect of a
measurement on uncertainty has been given by Peres~\cite{Peres1995}. Suppose the
random variable $X$ represents the location of a key, and prior knowledge has
established the following: the key is in my pocket with probability $p=0.9$, but if
it is not in my pocket, it can be in exactly 100 places with equal probability. The
random variable is thus actually composed of two {\em correlated} variables: the
pocket $P$ (with two states, {\em yes} and {\em no}), and the ``other" places $O$,
that has 100 states : $X=PO$. The entropy of $X$ is:
\begin{equation}
H(X)=H(O|P)+H(P)\;, \label{peres}
\end{equation}
where naturally $H(P)$ is my uncertainty about whether the key is in my pocket,
given by $H(P)=-0.1\ln 0.1+0.9\ln 0.9 \approx 0.325$, and $H(O|P)$ is the average
conditional entropy of the ``other" places, given I know whether or not the key is
in my pocket. So:
\begin{eqnarray}
H(O|P)&=& p\,H(O|P={\rm yes})+(1-p)\,H(O|P={\rm no}) \nonumber \\
&=&0.1\times 0 + 0.1\times \ln(100)\approx 0.4605\;,
\end{eqnarray}
since if the key is in my pocket, it is not in any of the 100 other places. Thus,
my uncertainty about the key location is $H(X)\approx0.7856$. Now, this type of
example is often used to claim that a measurement can sometimes {\em increase}
uncertainty, by nothing that, should I {\em not} find the key in my pocket ($P={\rm
no}$), my uncertainty is now $\ln 100\approx4.605$, much larger than 0.7856! But it
is in fact not $H(X)$ that has increased, since the new uncertainty is of course
just $H(X|P={\rm no})$, a {\em conditional} uncertainty. The entropy of $X$ was
changed only by reducing $H(P)$ in Eq.~(\ref{peres}) to zero (since the state of
the pocket will be known with certainty after the measurement), and therefore
\begin{equation}
H(X)\rightarrow H(O|P)\approx 0.4605\;.
\end{equation}
Thus, conditional entropies can increase or decrease due to a measurement, but the
unconditional entropy {\em must} decrease. This example is also instructive to
illustrate that in almost all cases, the entropy of random variables in physics is
going to be conditional on the state of other, measured variables. Indeed, {\em
subjectively} you sense that your uncertainty about the key's location has
increased after not finding it in your pocket, because {\em your} uncertainty has
become a conditional one after measurement. The fact that it has decreased {\em on
average} is irrelevant to you as an observer, because this may be the one and only
time you perform the measurement.

\subsection{Information and the Second Law}
\label{2ndlaw} Thermodynamics' second law is often regarded as one of physics' most
curious, because it appears to be intuitively correct while it cannot be derived
from first principles. I will take the position here that this is so because the
second law is usually formulated without giving sufficient heed to the notion that
conditional and unconditional entropies are fundamentally different, both from the
point of view of our intuition and from their mathematical structure. I have argued
above that thermodynamics can be viewed as a special case of information theory,
and I elaborate this point here.

The second law makes a prediction about the behavior of closed systems that evolve
from a non-equilibrium state towards an equilibrium state. In particular, the
second law predicts that the entropy of such a system will {\em almost always}
increase. The central observation about the inconsistency of this formulation lies
in recognizing that the second law describes {\em non-equilibrium dynamics} using
an equilibrium concept (namely Boltzmann-Gibbs entropy). Above, we have seen that
Shannon entropies, once we have chosen thermodynamical variables such as position
and momentum as those relevant to us, turn into Shannon-Gibbs entropies if
equilibrium distributions are used in Eq.~(\ref{therment}). But while a system
moves from non-equilibrium to equilibrium, we certainly cannot do this. Indeed, we
know that as a system equilibrates, conditional and unconditional entropies are
{\em not} equal. In order to correctly describe this, we have to use information
theory.

\subsubsection{Equilibration}
\begin{figure}
\vskip 0.5cm
\par
\centerline{\includegraphics[width=6cm,angle=-90]{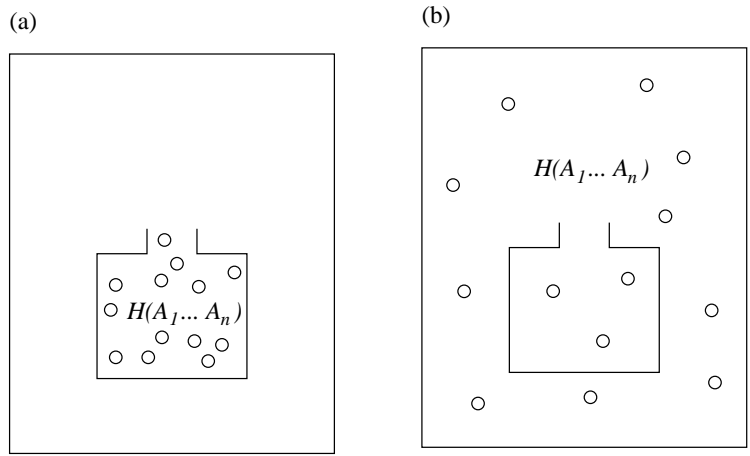}}\vskip 0cm
\caption{Diffusion of an ideal gas from a small into a larger container. (a) The
molecules with entropy $H(A_1\cdots A_n)$ occupy the smaller
  volume, and their correlation entropy is zero. (b) The molecules
  have escaped into the larger container, which increases the sum of
  the per-particle entropies and increases the correlation entropy
  commensurately such that the overall entropy remains unchanged.\label{equil}}
  \par
\end{figure}

Let us analyze the quintessential irreversible dynamics, the notorious ``perfume
bottle'' experiment, in which a diffusive substance (let's say, an ideal gas) is
allowed to escape from a small container into a larger one (see Fig.~\ref{equil}a).
Both the initial and the final state of the system can be described by equilibrium
equations; common wisdom however states that the entropy of the gas is {\em
increasing} during the process, reflecting the non-equilibrium dynamics.
I shall now show that this is not the case, by describing the gas in the smaller
container by a set of variables $A_1,\cdots,A_n$, one for each molecule. (What I
will show is that it is instead a {\em conditional} entropy that is increasing.)
The entropy $H(A_i)$ thus represents the entropy per molecule. The entire gas, on
the other hand, is described by the {\em joint entropy}
\begin{equation}
H_{\rm gas}=H(A_1\cdots A_n)\;, \label{joint}
\end{equation}
which can be much smaller than the sum of per-particle entropies because there are
strong {\em correlations} between the variables $A_i$. In information theory, such
correlations are described by information terms such as (\ref{info}). And as we
discussed above, they must vanish at equilibrium. The sum of per-particle
entropies, because it ignores correlations between subsystems, is just the standard
thermodynamical entropy $S_{eq}$
\begin{equation}
H(A_1\cdots A_n)\ll\sum_{i=1}^n H(A_i)=S_{eq}\;.\label{additive}
\end{equation}
The difference between $S_{eq}$ and (\ref{joint}) is given by the $n$-body
correlation entropy
\begin{equation}
H_{\rm corr}= \sum_{i=1}^n H(A_i)-H(A_1\cdots A_n) \;,\label{corr}
\end{equation}
which can be calculated in principle, but becomes cumbersome already for more than
three particles.

We see that in this description, the molecules after occupying the larger volume
cannot be independent of each other, as their locations are {\em in principle}
correlated (as they all used to occupy a smaller volume, see Fig.~\ref{equil}a). It
is true that once the molecules occupy the larger volume (Fig.~\ref{equil}b) the
observer has lost track of these correlations, and the second law characterizes
just how much information was lost. This statement, however, has nothing to do with
physics, but rather concerns an observer's capacity to make predictions. Indeed, these
correlations are not manifest in two-- or even three-body correlations, but are
complicated $n$-body correlations which imply that their positions are not
independent, but linked by the fact that they share initial conditions. This state
of affairs can be summarized by rewriting Eq.~(\ref{corr}):
\begin{equation}
H(A_1\cdots A_n) =\sum_{i=1}^n H(A_i)-H_{\rm corr}\;.
\end{equation}
We assume that before the molecules are allowed to escape, they are uncorrelated
with respect to each other: $H_{\rm corr}=0$, and the entire entropy is given by
the extensive sum of the per-molecule entropies. After expansion into the larger
volume, the standard entropy increases because of the increase in available phase
space, but this increase is balanced by an increase in the correlation entropy
$H_{\rm corr}$ in such a manner that the actual joint entropy of the gas, $H_{\rm
gas}$, remains unchanged.

Note that this description is not, strictly speaking, a {\rm redefinition} of
thermodynamical entropy. While in the standard theory, entropy is an {\em
extensive} (i.e., additive) quantity for uncorrelated systems, the concept of a
thermodynamical entropy in the absence of equilibrium distributions has been
formulated as the number of ways to realize a given set of occupation numbers of
states of the joint system (which gives rise to (\ref{entropy}) by use of
Stirling's approximation, see, e.g., \cite{Wannier1966}) and is thus fundamentally
{\em non-extensive}. Assuming the $A_i$ are uncorrelated reduces
$H(A_1\cdots A_n)$ to the extensive sum $\sum_{i=1}^{n}H(A_i)$, and thus to an
entropy proportional to the volume they inhabit. From a calculational point
of view the present formalism does not represent a great advantage in this case, as
the correlation entropy $H_{\rm corr}$ can only be obtained in special situations,
when only few-body correlations are important.

The examples of non-equilibrium processes treated here (measurement and
equilibration) suggest the following information-theoretical reformulation of the
second law:
\begin{quote}
\em In a thermodynamical equilibrium or non-equilibrium process, the
  unconditional (joint) entropy of a closed system remains a constant.
\end{quote}
Nothing can be said in principle about the conditional entropies involved (namely
the conditional entropy of the system given the state of the observer, or the
conditional entropy of the observer, given the state of the system), because they
can be increasing or decreasing. In a measurement, the conditional entropy
decreases (but the conditional entropy given a particular outcome can increase),
while during equilibration, the conditional entropy usually increases. That it can
sometimes {\em de}crease is acknowledged in the standard formulation of the second law by
the words ``almost always".  We recognize this as just one of these
rare fluctuations encountered in Section \ref{measexample}, where the entropy conditional on a particular outcome behaves counter-intuitively, while on average everything is as it should be.

The formulation of the second law given above directly reflects probability
conservation (in the sense of the Liouville theorem), and allows a quantitative
description of the amount by which either the conditional entropy is decreased in a
measurement, or the amount of per-particle entropy is increased in an equilibration
process.

\section{Quantum Theory}
In quantum mechanics, the concept of entropy translates very easily, but the
concept of information is thorny. John von Neumann introduced his eponymous quantum
mechanical entropy as early as 1927~\cite{vonNeumann27}, a full 21 years before
Shannon introduced its classical limit! In fact, it was von Neumann who suggested
to Shannon to call his formula entropy, simply because ``your uncertainty function
has been used in statistical mechanics under that name''~\cite{Tribus71}.

\subsection{Measurement}
In quantum mechanics, measurement plays a very prominent role, and is still
considered somewhat mysterious in many respects. The proper theory to describe
measurement dynamics in quantum physics, not surprisingly, is quantum information
theory. As in the classical theory, the uncertainty about a quantum system can only
be defined in terms of the detector states, which in quantum mechanics are a
discrete set of {\em eigenstates} of a measurement operator. The quantum system
itself is described by a wave function, given in terms of the quantum system's
eigenbasis, which may or may not be the same as the measurement device's basis.

For example, say we would like to ``measure an electron''. In this case, we may
mean that we would like to measure the position of an electron, whose wave function
is given by $\Psi(q)$, where $q$ is the coordinate of the electron. Further, let
the measurement device be characterized initially by its eigenfunction
$\phi_0(\xi)$, where $\xi$ may summarize the coordinates of the device. Before
measurement, i.e., before the electron interacts with the measurement device, the
system is described by the wave function
\begin{equation}
\Psi(q)\phi_0(\xi)\;.
\end{equation}
After the interaction, the wave function is a superposition of the eigenfunctions
of electron and measurement device
\begin{equation}
\sum_n \psi_n(q)\phi_n(\xi)\;. \label{sum}
\end{equation}
Following orthodox measurement theory, the classical nature of the measurement
apparatus implies that after measurement the ``pointer'' variable $\xi$ takes on a
well-defined value at each point in time; the wave function, as it turns out, is
thus {\em not} given by the entire sum in (\ref{sum}) but rather by the single term
\begin{equation}
 \psi_n(q)\phi_n(\xi)\;.\label{term}
\end{equation}
The wave function (\ref{sum}) is said to have collapsed to (\ref{term}).
\par
Let us now study what actually happens in such a measurement in detail. For ease of
notation, let us recast this problem into the language of state vectors instead.
The first stage of the measurement involves the interaction of the quantum system
$Q$ with the measurement device (or ``ancilla'') $A$. Both the quantum system and
the ancilla are fully determined by their state vector, yet, let us assume that the
state of $Q$ (described by state vector $|x\rangle$) is unknown whereas the state
of the ancilla is prepared in a special state $|0\rangle$, say. The state vector of
the combined system $|QA\rangle$ before measurement then is
\begin{equation}
|\Psi_{t=0}\rangle = |x\rangle|0\rangle \equiv |x,0\rangle\;.
\end{equation}
The von Neumann measurement is described by the unitary evolution of $QA$ via the
interaction Hamiltonian
\begin{equation}
\hat H = -\hat X_Q \hat P_A \;,\label{ham}
\end{equation}
operating on the product space of $Q$ and $A$. Here, $\hat X_Q$ is the observable
to be measured, and $\hat P_A$ the operator  {\it conjugate} to the degree of
freedom of $A$ that will reflect the result of the measurement. We now obtain for
the state vector $|QA\rangle$ after measurement (e.g., at $t=1$, putting $\hbar=1$)
\begin{equation}
|\Psi_{t=1}\rangle=e^{i\hat X_Q \hat P_A}|x,0\rangle =
 e^{ix\hat P_A}|x,0\rangle = |x,x\rangle \;. \label{entang}
\end{equation}
Thus, the pointer $A$ that previously pointed to zero now also points to the
position $x$ that $Q$ is in. This operation appears to be very much like the
classical measurement process Eq.~(\ref{cmeas}), but it turns out to be quite
different. In general, this unitary operation introduces quantum {\em
entanglement}, which is beyond the classical concept of correlations.

This becomes evident if we apply the unitary operation described above to an
initial quantum state which is in a quantum {\em superposition} of two states:
\begin{equation}
|\Psi_{t=0}\rangle = |x+y,0\rangle\;.
\end{equation}
Then, the linearity of quantum mechanics implies that
\begin{equation}
|\Psi_{t=1}\rangle = e^{i\hat X_Q\hat P_M} \biggl(|x,0\rangle+|y,0\rangle\biggr)=
|x,x\rangle+|y,y\rangle\label{entang1}\;.
\end{equation}
This state is very different from what we would expect in classical physics,
because $Q$ and $A$ are not just correlated (like, e.g., the state
$|x+y,x+y\rangle$ would be) but rather they are {\em quantum entangled}. They now
form {\em one} system that cannot be thought of as composite.

This nonseparability of a quantum system and the device measuring it is at the
heart of all quantum mysteries. Indeed, it is at the heart of {\em quantum
randomness}, the puzzling emergence of unpredictability in a theory that is
unitary, i.e., where all probabilities are conserved.  What is being asked here of
the measurement device, namely to describe the system $Q$, is logically impossible
because after entanglement the system has grown to $QA$. Thus, the detector is
being asked to describe a system that is {\em larger} (as measured by the possible number of
states) than the detector, and that includes the detector {\em itself}. This is
precisely the same predicament that befalls a computer program that is asked to determine
its own halting probability, in Turing's famous {\it Halting Problem} analogue~\cite{Turing1936} of
G\"odel's Incompleteness Theorem. Chaitin~\cite{Chaitin97} showed that the
self-referential nature of the question that is posed to the program gives rise to
randomness in pure Mathematics. A quantum measurement is self-referential in the
same manner, since the detector is asked to describe its own state, which is
logically impossible\footnote{The logical impossibility of describing one's own
state is intrinsically the same as that posed by the Cretan Paradox (Epimenides the
Cretan says ``All Cretans are liars.")}. Thus we see that quantum randomness has
mathematical (or rather logical) randomness at its very heart.

\subsection{von Neumann Entropy}
Because of this inherent uncertainty, measurements of a quantum system $A$ are then
described as {\it expectation values}, which are averages of an observable over the
system's {\it density matrix}, so that \be \la \hat O\ra = {\rm Tr} (\rho_A \hat
O)\;, \ee where $\hat O$ is an operator associated with the observable we would
like to measure, and $\rho_A$ is the density matrix of system $A$.  The latter is
obtained from the quantum wave function $\Psi_{QA}$ (for the combined system $QA$,
since neither $Q$ nor $A$ separately have a wave function after the entanglement
occurred) by {\it tracing out} the quantum system:
\begin{equation}
\rho_A = {\rm Tr}_Q |\Psi_{QA}\ra\la \Psi_{QA}|  \;. \label{parttrace}
\end{equation}
The partial trace represents an averaging over the states of the quantum system,
which after all is {\em not} being observed: we are looking at the measurement
device only. The uncertainty about quantum system $A$ can then be calculated simply
by von Neumann's entropy: \be S(\rho_A)= -\Tr\rho_A \log \rho_A \label{vnent}\;.
\ee
If $Q$ has been measured in $A$'s eigenbasis, then the density matrix $\rho_A$ is
diagonal, and von Neumann entropy turns into Shannon entropy, as we expect. Indeed,
this is precisely the classical limit, because entanglement does not happen under
these conditions.

Quantum Information Theory needs concepts such as conditional entropies and mutual
entropies. They can be defined in a straightforward manner~\cite{CerfAdami97}, but
their interpretation needs care.  For example, we can define a conditional entropy
in analogy to Shannon theory as 
\be
S(A|B)&=&S(AB)-S(B) \\
&=&-\Tr_{AB}(\rho_{AB}\log \rho_{AB}) +\Tr_B(\rho_B\log\rho_B)\;, \nonumber
\ee
where $S(AB)$ is the joint entropy of two systems $A$ and $B$. But can we write
this entropy in terms of a conditional density matrix, just as we were able to
write the conditional Shannon entropy in terms of a conditional probability? The
answer is yes and no: a definition in terms of a conditional density {\it operator}
$\rho_{A|B}$ exists~\cite{CerfAdami97,CerfAdami99}, but it is technically not a
density matrix (its trace is not equal to one), and the eigenvalues of this matrix
are very peculiar: they can {\em exceed} one (this is of course not possible for
probabilities). Indeed, they can exceed one only when the system is entangled. As a
consequence, quantum conditional entropies can be {\em
negative}~\cite{CerfAdami97}.

Even thornier is quantum mutual entropy. We can again define it simply in analogy
to (\ref{infosymm}) as \be S(A:B)=S(A)+S(B)-S(AB)\;, \ee but what does it mean? For
starters, this quantum mutual entropy can be twice as large as the entropy of any
of the subsystems, so $A$ and $B$ can share more quantum entropy then they even
have by themselves! Of course, this is due to the fact, again, that ``selves'' do
not exist anymore after entanglement. Also, in the classical theory, information,
that is, shared entropy, could be used to make predictions, and therefore to reduce
the uncertainty we have about the system that we share entropy with. But that's not
possible in quantum mechanics. If, for example, I measure the spin of a quantum
particle that is in an even superposition of its spin-up and spin-down state, my
measurement device will show me spin-up half the time, and spin-down half the time,
that is, my measurement device has an entropy of one bit. It can also be shown that
the shared entropy is {\it two bits}~\cite{CerfAdami97}. But this shared entropy
cannot be used to make predictions about the actual spin. Indeed, I still do not
know anything about it! On the other hand, it is possible, armed with my
measurement result, to make predictions about the state of {\em other} detectors
measuring the same spin. And even though all these detectors will agree about their
result, technically they agree about a random variable, not the actual state of the
spin they believe their measurement device to reflect~\cite{AdamiCerf99}. Indeed,
what else could they agree on, since the spin does not have a state? Only the
combined system with all the measurement devices that have ever interacted with it,
does.

Information, it turns out, is a concept that is altogether classical. {\it Quantum}
information, in hindsight, is therefore really a contradiction in terms. But that
does not mean that the entire field of quantum information theory is absurd.
Rather, what we mean by ``quantum information theory'' is the study of storage,
transmission, and manipulation of {\it qubits} (the quantum analogues of the usual
bit), which are quantum particles that can exist in superpositions of zero and one.
Indeed, the capacity of quantum channels to transmit classical information is
higher than any classical channel~\cite{AdamiCerf1997,Bennettetal2002}, for example, and
quantum bits can be used for super-fast computation~\cite{Shor94}.

The extension of Shannon's theory into the quantum regime not only throws new light
on the measurement problem, but it also helps in navigating the boundary between
classical and quantum physics. According to standard lore, quantum systems (meaning
systems described by a quantum wave function) ``become'' classical in the
macroscopic limit, that is, if the action unit associated with that system is much
larger than $\hbar$. Quantum information theory has thoroughly refuted this notion,
since we now know that macroscopic bodies can be entangled just as microscopic ones
can~\cite{julsgaard2001}. Instead, we realize that quantum systems appear to follow
the rules of classical mechanics if parts of their wave function are averaged over
[such as in Eq.~(\ref{parttrace})], that is, if the experimenter is not in total
control of all the degrees of freedom that make up the quantum system. Because
entanglement, once achieved, is not undone by the distance between entangled parts,
almost all systems will seem classical unless expressly prepared, and then
protected from interaction with uncontrollable quantum systems. Unprotected quantum
systems spread their state over many variables very quickly: a process known as
{\it decoherence} of the quantum state.

\section{Relativistic Theory}
Once convinced that information theory is a statistical theory about the relative
states of detectors in a physical world, it is clear that we must worry not only
about quantum detectors, but about moving ones as well. Einstein's special relativity
established an upper limit for the speed at which information can be transmitted without 
the need to cast this problem in an information-theoretic language. But in hindsight, it is clear that the impossibility of superluminal signaling could just as well have been the result of an analysis of the information transmission capacity of a communication channel involving detectors moving at constant speed with respect to each other. As a matter of fact, the capacity of an additive white noise Gaussian (AWNG) channel for information transmission for the case of moving observers just turns out to be~\cite{JarettCover1981}
\be
C=W\log(1+\alpha SNR)\;,
\ee
where $W$ is the bandwidth of the channel, $SNR$ is the signal-to-noise ratio, and $\alpha=\nu^\prime/\nu$ is the Doppler shift. As the relative velocity $\beta\rightarrow1$, 
$\alpha\rightarrow0$ and the communication capacity vanishes. 

Historically, however,
no-one seems to have worried about an ``information theory of moving bodies'', not
the least because such a theory had, or indeed has, little immediate relevance. (The above-mentioned reference~\cite{JarettCover1981} is essentially unknown in the literature.) A standard scenario of relativistic information theory would involve two random variables moving with respect to each
other. The question we may ask is whether and how relative motion is going to affect any
shared entropy between the variables. First, it is important to point out that
Shannon entropy is a {\em scalar}, and we therefore do not expect it to transform
under Lorentz transformations. This is also intuitively clear if we adopt the
``strict'' interpretation of entropy as being {\em unconditional} (and therefore
just equal to the logarithm of the number of degrees of freedom). On the other
hand, probability distributions (and the associated conditional entropies) could
conceivably change under Lorentz transformations. How is this possible given the
earlier statement that entropy is a scalar?

We can investigate this with a {\em gedankenexperiment} where the system under
consideration is an ideal gas, with particle velocities distributed according to an
arbitrary distribution. In order to define entropies, we have to agree on which
degrees of freedom we are interested in. Let us say that we only care about the two
components of the velocity of particles confined in the $x-y$-plane. Even at rest,
the mutual entropy between the particle velocity components $H(v_x:v_y)$ is
non-vanishing, due to the finiteness of the magnitude of $v$. A detailed
calculation~\cite{Gingrichpc} using continuous variable entropies of a uniform
distribution shows that, at rest
\begin{equation}
H(v_x:v_y)=\log(\pi/e) \label{bobeq}\;.
\end{equation}
The velocity distribution, on the other hand, will surely change under Lorentz
transformations in, say, the $x$-direction, because the components are affected
differently by the boost. In particular, it can be shown that the mutual entropy
between $v_x$ and $v_y$ will {\em rise} monotonically from $\log(\pi/e)$, and tend
to a constant value as the boost-velocity $\beta \rightarrow1$~\cite{Gingrichpc}.
But of course, $\beta$ is just another variable characterizing the moving system,
and if this is known precisely, then we ought to be able to recover
Eq.~(\ref{bobeq}), and the apparent change in information is due entirely to a
reduction in the uncertainty $H(v_x)$. Similar conclusions can be reached if the
Maxwell distribution is substituted for the uniform one. This example shows that in
information theory, even if the entire system's entropy does not change under
Lorentz transformations, the entropies of subsystems, and therefore also
information, can.

While a full theory of relativistic information does not exist, pieces of such a
theory can be found when digging through the literature, For example, relativistic
thermodynamics is a limiting case of relativistic information theory, simply
because as we have seen above, thermodynamical entropy is a limiting case of
Shannon entropy. But unlike in the case constructed above, we do not have the
freedom to choose our variables in thermodynamics. Hence, the invariance of entropy
under Lorentz transformations is assured via Liouville's theorem, because the
latter guarantees that the phase-space volume occupied by a system is invariant.
Yet, relativistic thermodynamics is an odd theory, not the least because it is
intrinsically inconsistent: the concept of equilibrium becomes dubious. In
thermodynamics, equilibrium is defined as a state where all relative motion between
the subsystems of an ensemble has ceased. Therefore, a joint system where one part
moves with a constant velocity with respect to the other cannot be at equilibrium,
and relativistic information theory has to be used instead.

One of the few questions of immediate relevance that relativistic thermodynamics
has been able to answer is how the temperature of an isolated system will appear
from a moving observer. Of course, temperature itself is an equilibrium concept and
therefore care must be taken in framing this question~\cite{Aldrovandi92}. Indeed,
both Einstein and Planck~\cite{Old} tackled the question of how to
Lorentz-transform temperature, with different results. The
controversy~\cite{Ottarz6364} can be resolved by realizing that no such
transformation law can in fact exist~\cite{Landsberg96}, as the usual temperature
(the parameter associated with the Planckian blackbody spectrum) becomes
direction-dependent if measured with a detector moving with velocity $\beta=v/c$
and oriented at an angle $\theta^\prime$ with respect to the
radiation~\cite{Pauli21,Peebles68} \be T^\prime=\frac{T\sqrt{1-\beta^2}}{1-\beta
\cos\theta^\prime}\;. \ee In other words, an ensemble that is thermal in the rest
frame is non-thermal in a moving frame, and in particular cannot represent a
standard heat bath because it will be {\em non-isotropic}.

\section{Relativistic Quantum Theory}
While macroscopic quantities like temperature lose their meaning in relativity,
microscopic descriptions in terms of probability distributions clearly still make
sense. But in a quantum theory, these probability distributions are obtained from 
quantum measurements specified by local operators, and the space-time relationship between the detectors implementing these operators becomes important.
For example, certain measurements on a joint (i.e., composite) system may require communication between parties, while  certain others are impossible 
{\em even though} they do not require communication~\cite{Beckman01}. In general, 
a relativistic theory of quantum information needs to pay close attention to the behavior of the von Neumann  entropy under Lorentz transformation, and how such entropies are being reduced by measurement.

\subsection{Boosting Quantum Entropy}
The entropy of a qubit (which we take here for simplicity to be a spin-1/2
particle) with wave function
\be
|\Psi\ra=\frac1{\sqrt{|a|^2+|b|^2}}\biggl(a|\uparrow\ra+b|\downarrow\ra\biggr)
\;,\label{qubit} 
\ee 
($a$ and $b$ are complex numbers), can be written in terms of
its density matrix $\rho=|\Psi\ra\la \Psi|$ as \be S(\rho)=-\Tr(\rho\log \rho)\;.
\label{qubitent} \ee A wave function is by definition a completely known state
(called a ``pure state''), because the wave function is a complete description of a
quantum system. As a consequence, (\ref{qubitent}) vanishes: we have no uncertainty
about this quantum system. As we have seen earlier, it is when that wave function
interacts with uncontrolled degrees of freedom that mixed states arise. And indeed,
just by boosting a qubit, such mixing will arise~\cite{Peres02}. The reason is not
difficult to understand.  The wave function (\ref{qubit}), even though I have just
stated that it completely describes the system, in fact only completely describes
the {\em spin} degree of freedom! Just as we saw in the earlier discussion about
the classical theory, there may always be other degrees of freedom that our
measurement device (here, a spin-polarization detector) cannot resolve. Because we
are dealing with particles, ultimately we have to consider their {\it momenta}. A
more complete description of the qubit state then is 
\be
|\Psi\ra = |\sigma\ra\times |\vec p\ra\;, \label{psqubit}
\ee
where $\sigma$ stands for the
spin-variable, and $\vec p$ is the particle's momentum. Note that the momentum wave
function $|\vec p\ra$ is in a product state with the spin wave function
$|\sigma\ra$. This means that both spin and momentum have their own state, they are
{\it unmixed}. But as is taught in every first-year quantum mechanics course, such
momentum wave functions (plane waves with perfectly well-defined momentum $\vec p$)
do not actually exist; in reality, they are wave packets with a momentum {\em
distribution} $f(\vec p)$, which we may take to be Gaussian. If the system is at
rest, the momentum wave function does not affect the entropy of (\ref{psqubit}),
because it is a product.

What happens if the particle is boosted? The spin and momentum degrees {\em do}
mix, which we should have expected because Lorentz transformations always imply
frame rotations as well as changes in linear velocity. The product wave function
(\ref{psqubit}) then turns into \be |\Psi\ra\longrightarrow \sum_\sigma \int f(\vec
p) |\sigma,\vec p\ra d\vec p\;, \ee which is a state where spin-degrees of freedom
and momentum degrees of freedom are entangled. But our spin-polarization detector
is insensitive to momentum! Then we have no choice but to average over the
momentum, which gives rise to a spin density matrix that is mixed, \be \rho_\sigma
= \Tr_{\vec p} (\rho_{\sigma \vec p})\;, \ee and which consequently has positive
entropy. Note, however, that the entropy of the joint spin-momentum density matrix
remains unchanged, at zero. Note also that if the momentum of the particle was
truly perfectly known from the outset, i.e., a plane wave $|\vec p\ra$, mixing
would also not take place~\cite{AlsingMilburn02}.

While the preceding analysis clearly shows what happens to the quantum entropy of a
spin-1/2 particle under Lorentz transformations (a similar analysis can be done for
photons~\cite{PeresTerno03a}), what is most interesting in quantum information
theory is the entanglement {\em between} systems. While some aspects of
entanglement are captured by quantum entropies~\cite{BDSW96} and the spectrum of
the conditional density operator~\cite{CerfAdami99}, quantifying entanglement is a
surprisingly hard problem, currently without a perfect solution. However, some good
measures exist, in particular for the entanglement between two-level systems
(qubits) and three-level systems.

\subsection{Boosting Quantum Entanglement}
If we wish to understand what happens to the entanglement between two spin-1/2
particles, say, we have to keep track of four variables: the spin states
$|\sigma\ra$ and $|\lambda\ra$ and the momentum states $|\vec p\ra$ and $|\vec
q\ra$. A Lorentz transformation on the joint state of this two-particle system will
mix spins and momenta just as in the previous example. In fact, it is known that
this type of mixing will affect the entanglement between pairs of particles that
are used to test the violation of Bell inequalities, for
example~\cite{Czachor1997}. In order to investigate this effect from the point of
view of quantum information theory, we need to study the behavior of an
entanglement measure under Lorentz boosts.

A good measure for the entanglement of mixed states, i.e., states that are not pure
such as (\ref{psqubit}), is the so-called {\it concurrence}, introduced by
Wootters~\cite{Wootters98}. This concurrence $C(\rho_{AB})$ can be calculated for a
density matrix $\rho_{AB}$ that describes two subsystems $A$ and $B$ of a larger
system, and quantifies the entanglement {\em between} $A$ and $B$. For our
purposes, we will be interested in the entanglement between the spins $\sigma$ and
$\lambda$ of our pair. The concurrence is unity if two degrees of freedom are
perfectly entangled, and vanishes if no entanglement is present.

In order to do this calculation we first have to specify our initial state. We take
this to be a state with spin and momentum wave function in a product, but where the
spin-degrees of freedom are perfectly entangled in a so-called Bell state: 
\be
|\sigma,\lambda\ra = \frac1{\sqrt{2}}\left(|\uparrow,\downarrow\ra -
|\downarrow,\uparrow\ra\right)\;. \label{bell}
\ee
Of course, such states have concurrence $C(\rho_{\sigma\lambda})=1$. We now apply a Lorentz
boost to this joint state, i.e., we move our spin-polarization detector with speed
$\beta=v/c$ with respect to this pair (or, equivalently, we move the pair with
respect to the detector). If the momentum degrees of freedom of the particles at
the outset are Gaussian distributions unentangled with each other and the spins,
the Lorentz boost will entangle them, and the concurrence {\em between the spins} will
drop~\cite{GingrichAdami02}. How much it drops depends on the ratio between the
spread of the momentum distribution $\sigma_r$ (not to be confused with the spin
$\sigma$) and the particle's mass $m$. In Fig.~\ref{conc} below, the concurrence is
displayed for two different such ratios, as a function of the {\it rapidity} $\xi$.
The rapidity $\xi$ is just a transformed velocity: $\xi = \sinh\beta$, such
that $\xi\rightarrow\infty$ as $\beta\rightarrow1$. We can see that if the ratio is
not too large, the concurrence will drop but not disappear altogether. But if the
momentum spread is large compared to the mass, all entanglement can be lost.

\begin{figure}[htb]
\begin{center}
\includegraphics[width=7.5cm,angle=90]{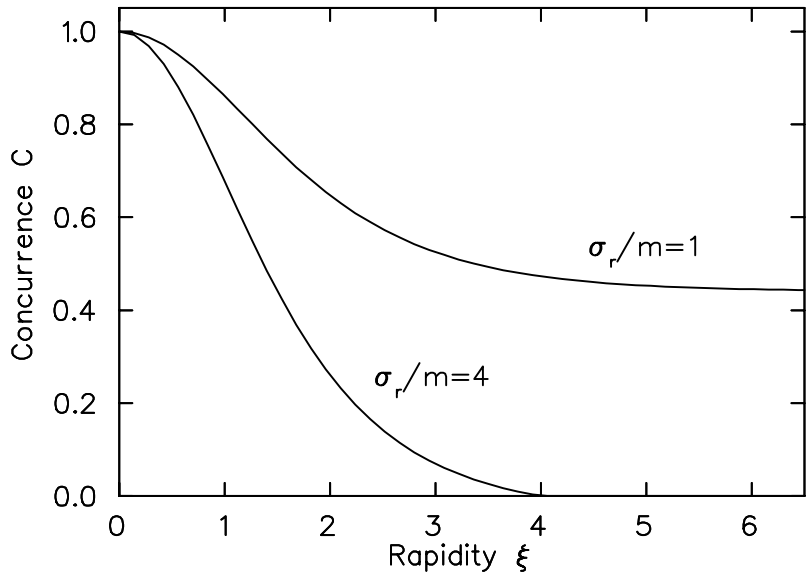}
\end{center}
\caption{Spin-concurrence as a function of rapidity, for an initial Bell state with
momenta in a product Gaussian.  Data is shown for $\sigma_r / m = 1$ and $\sigma_r
/ m = 4$ (from Ref.\protect\cite{GingrichAdami02})\label{conc}.}
\end{figure}

\begin{figure}[htb]
\begin{center}
\includegraphics[width=6.5cm,angle=0]{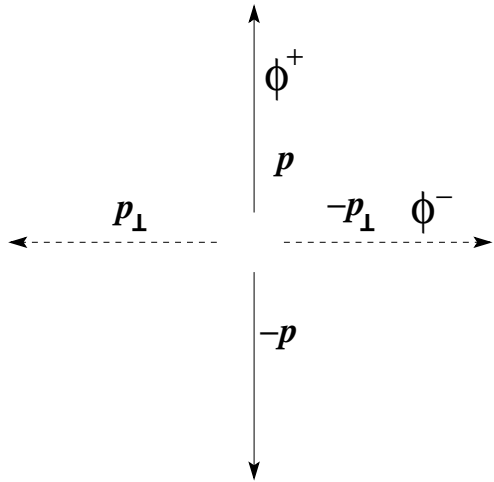}
\end{center}
\caption{Superposition of Bell-states $\Phi^+$ and $\Phi^-$ at right angles, with
the particle pair moving in opposite directions.}. \label{fig2}
\end{figure}

Let us consider instead a state that is unentangled in spins, but fully entangled
in momenta. I depict such a wave function in Fig.~\ref{fig2}, where a pair is in a
superposition of two states, one moving in opposite directions with momentum $\vec
p_\bot$ in a relative spin state $\Phi^-$ (this is one of the four Bell
spin-entangled states, Eq.~(\ref{bell})), and one moving in a plane in opposite
orthogonal directions with momentum $p$, in a relative spin-state $\Phi^+$. It can
be shown that if observed at rest, the spins are actually unentangled. But when
boosted to rapidity $\xi$, the concurrence  {\em
increases}~\cite{GingrichAdami02}, as for this state (choosing $m=1$)
\begin{equation}
C(\rho_{A B}) = \frac{p^2 (\cosh^2 (\xi) - 1)}{(\sqrt{1 + p^2} \cosh
  (\xi) + 1)^2}\;.
\end{equation}

A similar analysis can be performed for pairs of entangled photons, even though the
physics is quite different~\cite{GingrichBergouAdami03}. First of all, photons are
massless and their quantum degree of freedom is the photon-polarization. The
masslessness of the photons makes the analysis a bit tricky, because issues of
gauge invariance enter into the picture, and as they all move with constant
velocity (light speed), there cannot be a spread in momentum as in the massive
case. Nevertheless, Lorentz transformation laws acting on polarization vectors can
be identified, and an analysis similar to the one described above can be carried
through. The difference is that the entangling effect of the Lorentz boost is now
entirely due to the spread in momentum {\em direction} between the two entangled
photon beams. This implies first of all that fully-entangled photon polarizations
cannot exist, even for detectors at rest, and second that existing entanglement can either be
decreased or increased, depending on the angle with which the pair is boosted (with
respect to the angle set by the entangled pair), and the
rapidity~\cite{GingrichBergouAdami03}.

\section{Information in Accelerated Frames and Curved Space Time}
Relativistic quantum information theory is a growing field~\cite{PeresTerno03b} that has naturally engendered questions about the relative state of detectors in {\em non-inertial} frames. Accelerated detectors introduce a new twist to 
information theory: whether or not a detector registers depends on its state of motion, that is, even presence or absence can become relative! To some extent this is not a completely unfamiliar situation. We are used to radiation being emitted from accelerated charges, and from detectors moving through a medium with changing index of refraction~\cite{FultonRohrlich1960,FrolovGinzburg1986}. In such cases, the state of the detector depends on {\em which vacuum} it perceives. For all {\em inertial} measurement devices, all vacua are equivalent because they are invariant under the Poincar\'e group. Yet this invariance doesn't hold for accelerated observers or detectors in strong fields~\cite{BirrellDavies1982}, and no particular vacuum state is singled out. Indeed, different vacua can be defined using Bogoliubov transformations to transform one set of creation/annihilation operators into another, and in principle none would be preferred. How do we then make sense of the physical world, which requires that all observers agree about the result of measurements? Usually, the agreed-upon vacuum is the one where absence of particles is perceived by {\em all inertial observers}, for example in the remote past or future. This is also the approach taken below when we consider black hole formation and evaporation, where 
quantum states $|k\ra_{\rm in}$ and $|k\ra_{\rm out}$ refer to in- and out-states in past and future infinity. In the meantime, let us discuss briefly the relative state of non-inertial detectors. 

That something interesting must happen
to entropies in non-inertial frames is immediately clear from the Unruh
effect~\cite{Fulling1973,Davies1975,Unruh1976}. The Unruh effect is perhaps the
most important clue to our understanding of quantum field theory in curved space
time, which is still quite incomplete. Accelerated observers perceive a
vacuum quite different from that apparent to a non-accelerated observer: they find
themselves surrounded by thermal photons of temperature $T_U=\frac{\hbar a}{2\pi
c}$ (the Davies-Unruh temperature), where $a$ is the observer's acceleration, and
$c$ is the speed of light. If we were to calculate the entropy of a particle in the
inertial vs. the non-inertial frame, the absence or presence of the Unruh radiation
implies that they would be different. In other words, standard thermodynamic or
von Neumann entropies do not transform covariantly under general co-ordinate
transformations, that is, they are not scalars. Again this should not be surprising, because positive von Neumann entropies only occur if part of an entangled state is averaged over. Only the entropy of the pure state is invariant (it vanishes). 
There are immediate consequences for standard quantum information
protocols such as quantum teleportation: while the required resource (and entangled
pair) can be used to teleport one qubit perfectly in an inertial frame, the
appearance of Unruh radiation dilutes the entanglement such that the fidelity of
teleportation is reduced~\cite{Alsingetal2003}. There are also consequences when 
positive entropies are forced upon us simply because certain parts of spacetime are inaccessible to measurements. Such is the case beyond black-hole event horizons.

\subsection{Black Hole Information Paradox}
Ever
since the discovery of Hawking radiation~\cite{Hawking1975} we are faced with what is
known as the {\em black hole information paradox}~\cite{Preskill}. The paradox can
be summarized as follows. According to standard theory, a black hole (without charge or angular momentum) can be
described by an entropy~\cite{Bekenstein1973} that is determined entirely in terms of its mass $M$ (in units where $\hbar=c=G=k=1$): 
\be
 S_{BH}=4\pi M^2\;. \label{bhent}
\ee 
From the point of view adopted in this review, this formula implies that black holes are very peculiar objects. They have an entropy given by one quarter of the surface area of the black hole event horizon ($r=2M$ in these units), which we have to associate with an {\em uncertainty} in classical information theory. Yet, it seems we cannot learn anything about the black hole, because we cannot measure its states. Fortunately, quantum mechanics comes to the rescue here.
Classically, a black hole has zero temperature (because it does not radiate), but a quantum treatment shows that vacuum fluctuations near the event horizon cause the black hole to radiate like a black body at a temperature $T_H=(8\pi M)^{-1}$, the Hawking temperature. In a sense, the black hole polarizes the vacuum around it, causing spontaneous emission of radiation, and a {\em reduction} of entropy. The alert reader should make a mental note at this point, because I have argued earlier that, strictly speaking, only {\em conditional} entropies can decrease. Could it be that the black hole entropy is in fact a conditional entropy? Before we enter this discussion, I should summarize an apparently very alarming state of affairs in black hole physics.

If a state that is
fully known is absorbed by the black hole, i.e, if it disappears behind the event
horizon, it appears as if the information about the identity of the state is lost. Even worse, after a long time, the black hole will have evaporated entirely into thermal (Hawking) radiation, and the
information appears not only to be irretrievable, but {\em destroyed}. Indeed, it would appear that black holes have the capability to turn pure states into
mixed stated {\em without} disregarding parts of the wave function. Such a state of
affairs is not only paradoxical, it is in fact completely out of the question
within the standard model of physics.

It has been argued that this paradox stems from our incomplete understanding of 
quantum gravity. For example, in the semiclassical framework (in which Hawking's calculation was carried out~\cite{Hawking1975}), the space time metric remains unquantized, and is treated instead as a classical background field. 
A consistent treatment instead would allow particle degrees of freedom to be entangled with the metric, creating quantum mechanical uncertainty (see~\cite{KokYurtsever2003} for a calculation of the decoherence of a qubit in an orbit around a Scharzschild black hole due to entanglement with the metric). But it is difficult to conceive that this effect has a significant impact on black hole dynamics. Indeed, gravitational fields are weak up until the black hole is of the order of the Planck mass, when backreaction effects on the space time metric presumably become important~\cite{Wald1994}. But at this point, all the ``damage" has already been done, because it is difficult to imagine that 
lost information can be recovered from a Planck-sized object. (An enormous amount of entropy would have to be emitted instantaneously by a black hole of size $L_{\rm Planck} \approx 10^{-33}$ cm, see Ref.~\cite{Wald1994}, p.\ 184). Instead, we should look at 
the treatment of black holes within classical {\em equilibrium} thermodynamics as the culprit for the paradox. 

Black holes have negative heat capacity (they become hotter the more energy they radiate away), and therefore can {\em
never} be at equilibrium with a surrounding (infinite) heat bath. As we have seen, the concept of information {\em itself} is a non-equilibrium one (because information implies correlations). Moreover, quantum correlations (in the form of entanglement) can exist 
over arbitrary distances {\em and} across event horizons, because entanglement does not imply signaling. Thus, a black hole can be entangled with surrounding radiation even though the two are causally disconnected. These considerations make it likely that a quantum information-theoretic treatment of black hole dynamics could resolve the paradox without appealing to a consistent theory of quantum gravity. 
In the following, I outline just such  a scenario.

First, we must recognize that a positive entropy in quantum mechanics implies that the black hole is described by a density matrix of the form
\be
\rho_{\rm BH} = \sum_i p_i \rho_i\;
\ee\
where $\rho_i = |i\ra\la i|$ are obtained from black hole eigenvectors. Such a mixed state can always be written in terms of a pure wavefunction via a Schmidt decomposition:
\be
\rho_{\rm BH} = \sum_i \sqrt{p_i}|\psi_i\ra_{\rm BH}\ra |i\ra_{\rm R}\;,
\ee
where the $|i\ra_{\rm R}$ are the eigenstates of a ``reference" system. While such a Schmidt decomposition is always possible mathematically, what physics does it correspond to? 
If there is a physical mechanism that associates states $|i\ra_{\rm R}$  {\em outside} of the event horizon with states $|\psi_i\ra_{\rm BH}$ within it, then perhaps information entering a black hole will leave some signature outside it. As it turns out, there is indeed such a physical mechanism, as I now show.

Black hole evaporation is, as mentioned earlier, due to a quantum effect that has no analogue in classical physics: the spontaneous decay of the vacuum. Quite literally, the gravitational field surrounding the black hole polarizes the vacuum so as to create virtual particle-antiparticle states. If one member of such a pair enters the event horizon while the other goes off to infinity, the black hole itself will have to provide the energy to convert the virtual particles to real ones, i.e., to put them on their mass shell. This process reduces the black hole's mass, and thus, via Eq.~(\ref{bhent}), its entropy. Within quantum information theory, we write the black hole entropy in terms of the von Neumann entropy
\be
S(\rho_{\rm BH}) = - \tr \rho_{\rm BH}\log \rho_{\rm BH}\;,
\ee
and the reduction in entropy can be understood in terms of the {\em removal} of positive energy modes due to the absorption of negative energy modes. As pointed out earlier, such a description only makes sense if we consider {\em aymptotic} modes. Then, a flux of particles of positive energy at $t\rightarrow\infty$ must correspond to a flux of negative energy of equal magnitude into the black hole. In a sense, it is our detectors in the future that are allowing us to make predictions in the past. 

In  this picture then, we can describe particle absorption and emission from a black hole with a standard interaction Hamiltionian, so that for an incoming particle in mode $k$ incident on a black hole with wavefunction $|\psi\ra_{\rm Q}$ (the index Q will be used to label the black-hole Hilbert space throughout this section) we have the unitary evolution
\be
U|k\ra_{\rm M}|\psi\ra_{\rm Q}|0\ra_{\rm R}&=& \lim_{t\to\infty}e^{i\int_{-\infty}^\infty H_{\rm int}dt}|k\ra_{\rm M}|\psi\ra_{\rm Q}|0\ra_{\rm R}\nonumber \\
&=& |k\ra_{\rm M}|\psi\ra_{\rm Q}|0\ra_{\rm R} + \alpha_k|0\ra_{\rm M}|\psi_k\ra_{\rm Q} |0\ra_{\rm R} + \beta_k|k\ra_{\rm M}|\psi_{-k}\ra_{\rm Q} |k\ra_{\rm R}\;. \nonumber \\ \label{stimul}
\ee
Here, we defined black hole wavefunctions that have either absorbed or emitted a mode $|k\ra$ using the ``ladder operators" $\sigma_k^+$ and $\sigma_k^-$
\be
|\psi_{\pm k}\ra = \sigma_k^{\pm}|\psi\ra\;,
\ee
and introduced a Fock space for {\em stimulated} radiation, with ground state $|0\ra_{\rm R}$.
The coefficients $\alpha_k$ and $\beta_k$ are related to the Einstein coefficients for emission and absorption~\cite{BekensteinMeisels1977}, so that $\beta_k=\alpha_ke^{-i\omega_k/T_{\rm H}}$. Thus, we see that beyond spontaneous emission and absorption, stimulated emission plays a crucial role in black hole dynamics~\cite{Wald1976,AudretschMueller1992,Schiffer1993,MuellerLousto1994}. Indeed, it is the key to the purification of the black hole density matrix. If we ignore the elastic scattering term in Eq.~(\ref{stimul}) (it does not affect the entanglement), we see that the accretion of $n$ modes onto a pure black hole state $|\psi\ra$ gives rise to the entangled wavefunction
\be
|\Psi\ra_{\rm QMR}=\sum_{i=1}^{2^n}\sqrt{p_i}|\psi_i\ra_{\rm Q}|i\ra_{MR}\;, \label{total}
\ee
as promised. Information about the identity of the particle modes, i.e., the basis states of M given by the labels attached to each incoming mode $|k_i\ra_{\rm M}$, are encoded in the stimulated radiation states R in such a manner that they are perfectly correlated: $I=S(M:R)$. Because M and R share the same basis system, this von Neumann mutual entropy actually reduces to a Shannon information, and we can say that M plays the role of a 
{\em preparer} of the quantum states $|k_i\ra_{\rm R}$. The problem of understanding the fate of information in black hole evaporation now just reduces to a problem in quantum channel theory (namely the transmission of classical information through an entanglement-assisted channel~\cite{AdamiCerf1997,Bennettetal2002,Holevo2002}), because spontaneous emission of particles from state (\ref{total}) creates a noisy quantum channel for the classical information $I$. 

Because of the initial entanglement between the black hole and the stimulated radiation, the final state after spontaneous emission will be an entangled {\em pure} state between Q, the joint system MR, and the Hawking radiation. As the states MR are not being measured, we need to trace them out to consider the joint state of black hole and Hawking radiation. In particular, the Hawking radiation will appear completely thermal with temperature $T_{\rm H}$. And as long as no additional particles accrete, the wavefunction (\ref{total}) ensures that the entropy of the MR system (note that $S(MR)=S(M)=S(R)$) is {\em always equal} to the joint entropy of black hole and Hawking radiation. This implies that we ought to be able to reformulate the second law of black hole thermodynamics in a manner very similar to the modification introduced in section \ref{2ndlaw}. This second law~\cite{Bekenstein1973} (see also~\cite{FrolovPage1993} for a derivation in terms of quantum entanglement)  states that the sum of black hole entropy and surrounding matter/radiation (thermodynamical)  entropy can never decrease:
\be
dS_{\rm tot} = d(S_{\rm BH} + S_{\rm therm})\geq0\;. \label{2law}
\ee
But in information theory we can write {\em equalities} because we do not have to ignore correlations, i.e., information entropies. According to the scenario above, we can thus state that

{\em In black hole evaporation, the joint entropy of black hole and Hawking radiation, as well as the joint entropy of black hole, radiation, and the infalling matter distribution, remain a constant.} 

As a corollary, we note that the only entropy that can decrease in such a process is indeed the quantum entropy of the black hole given the outgoing radiation, as we suspected earlier. A detailed description of black hole formation and evaporation in a quantum information-theoretic setting is beyond the scope of this review, and will appear elsewhere~\cite{AdamiVerSteeg2004}.

\subsection{Entropies in Curved Space Time}
The treatment of black hole dynamics outlined above, while suggestive, did not fully use the formalism of quantum field theory in curved space time. For example,
entropies were calculated in terms of quantum mechanical wavefunctions, and in general 
our information degrees of freedom were particles with particular quantum numbers. But as we saw earlier, a description of detectors in non-inertial frames is inconsistent because whether or not a detector fires depends on its acceleration. Indeed, the particle concept itself is suspect in this context, and instead we should use an approach based entirely on quantum fields and their fluctuations. Such a formalism is preferred also because quantum field theory guarantees that observables interact in a manner compatible with the causal structure of space-time. Thus, in order to
consistently define quantum entropies in curved space-time, we must define them within quantum field theory. To close this review, I briefly speculate about such an approach.

The first steps toward such a theory involve defining quantum fields over a
manifold separated into an accessible and an inaccessible region. This division
will occur along a world-line, and we shall say that the ``inside'' variables are
accessible to me as an observer, while the outside ones are not. Note that the
inaccessibility can be due either to causality, or due to an event horizon. Both
cases can be treated within the same formalism (and indeed the derivation of the Unruh and Hawking effect are very similar for this reason). 
States in the inaccessible region
have to be averaged over, since states that differ only in the outside region are
unresolvable. Let me denote the inside region by $R$, while the entire state is
defined on $E$. We can now define a set of commuting variables $X$ that can be
divided into $X_{\rm in}$ and $X_{\rm out}$. By taking matrix elements of the
density matrix of the entire system 
\be 
\rho=|E\ra\la E| 
\ee 
with the complete set
of variables ($X_{\rm in},X_{\rm out}$), we can construct the inside density matrix
(defined on $R$) as 
\be
 \rho_{\rm in}=\Tr_{X_{\rm out}}(\rho_{X_{\rm in}X_{\rm
out}})\;.
\ee
This allows me to define the {\em geometric
entropy}~\cite{Bombellietal1986,CallanWilczek1994,Holzhey94} of a state on $E$ for an observer restricted to $R$ 
\be
S_{\rm geom}= -\Tr(\rho_{\rm in}\log \rho_{\rm in})\;, \label{geom} \ee 
where the trace is performed using the inside variables only. For quantum fields with equiprobable modes, we can see this expression as giving the logarithm of the number of states in the inaccessible (i.e., ``out") region that are {\em consistent} with measurements restricted to the ``in'' region~\cite{Holzhey94}.
Writing down such an expression, however, is just the beginning. 

As with most quantities in quantum field
theory, the geometric entropy (\ref{geom}) is divergent and needs to be renormalized. Rather than
being an inconvenience, this is precisely what we should have expected: after all,
we began this review by insisting that entropies only make sense when discussed in
terms of the possible measurements that can be made of the system. This is, of
course, precisely the role of renormalization in quantum field theory. Quantum
entropies can be renormalized via a number of methods, either using Hawking's zeta
function regularization procedure~\cite{Hawking77} or by the ``replica trick'',
writing \be S_{\rm geom} = -\left[\frac{d}{dn}\Tr(\rho_{\rm in}^n)\right]_{n=1}\;,
\ee and then writing $dS(n)$ in terms of the expectation value of the stress
tensor. A thorough application of this program should reveal components of the
geometric entropy due entirely to the curvature of space-time, components that vanish in the
flat-space limit. Furthermore, the geometric entropy can be used to write equations
relating the entropy of the inside and the outside space-time regions, as \be
S(E)=S(\rho_{\rm in, out}) = S(\rho_{\rm in}) + S(\rho_{\rm out}|\rho_{\rm in})\;.
\ee
A thorough application of this program, with appropriate renormalization of both ultraviolet and infrared divergencies, should finally yield an origin of the mysterious Bekenstein entropy fully in accord with information theory. First steps in this direction have indeed been taken very recently by Terno~\cite{Terno2004}, who studied the transformation properties of geometric entropy, and found that $S_{\rm geom}$ is not a scalar under Lorentz transformation, while the Bekenstein-Hawking entropy is. Clearly, 
we are still not close to a full quantum field-theoretic description of information in arbitrary 
space-times, but it would appear that the necessary tools are available.

\section{Summary}
Entropy and information are statistical quantities describing an observer's
capability to predict the outcome of the measurement of a physical system. Once
couched in those terms, information theory can be examined in all physically
relevant limits, such as quantum, relativistic, and gravitational. Information
theory is a non-equilibrium theory of statistical processes, and should be used
under circumstances (such as measurement, non-equilibrium phase transitions,
etc.) where an equilibrium approach is inappropriate. Because an observer's capability
to make predictions (quantified by entropy) is not a characteristic of the object
the predictions apply to, it does not have to follow the same physical laws (such
as reversibility) as that befitting the objects. Thus, the arrow of time implied by
the loss of information under standard time-evolution is even less mysterious than
the second law of thermodynamics, which is just a consequence of the former.

In time, a fully relativistic theory of quantum information, defined for quantum fields on curved
space-time, should allow us to tackle a number of problems in cosmology and other
areas that have as yet resisted a consistent treatment. These developments, I have
no doubt, would have made Shannon proud.

\ack I am grateful to N.~J. Cerf for years of very fruitful collaboration in quantum
information theory, as well as to R.~M. Gingrich and A.~J. Bergou for their joint
efforts in the relativistic theory, and G.~L. Ver Steeg for collaboration on black-hole dynamics. I would also like to acknowledge crucial
discussions on entropy, information, and black holes, with P. Cheeseman, J.P.
Dowling, and U. Yurtsever in particular, and the Quantum Technologies Group at JPL in general. This work was carried out in part at the Jet Propulsion
Laboratory (California Institute of Technology) under a contract with the National
Aeronautics and Space Administration, with support from the Army Research Office's grant \# DAAD19-03-1-0207.


\section*{References}



\begin{thebibliography}{99}
\bibitem{Shannon48}C. Shannon, A mathematical theory of communication.
{\it Bell System Tech. Jour.} {\bf 27}, 379-423 (1948) {\it ibid}, 623-656.
\bibitem{Landauer91}R. Landauer, Information is physical. {\it
Phys. Today} {\bf 44}, 23-29 (1991).
\bibitem{CoverThomas}T.~M. Cover and J.~A. Thomas, {\it Elements of
Information Theory}. (Wiley, New York, 1991).
\bibitem{Jaynes1957}E.~T. Jaynes, Information theory and statistical physics. {\it
Phys. Rev.} {\bf 106}, 620-630 (1957).
\bibitem{Peres1995}A. Peres, {\it Quantum Theory: Concepts and Methods}. (Kluwer
Academic, Dordrecht, 1995).
\bibitem{Wannier1966}G.~H. Wannier, {\it Statistical Physics}. (Wiley, New
  York, 1966).
\bibitem{vonNeumann27}J. von Neumann, Thermodynamik
quantenmechanischer Gesamtheiten. {\it G\"ott. Nach.} {\bf 1}, 272-291 (1927).
\bibitem{Tribus71}M. Tribus and E.~C. McIrvine, Energy and information.
{\it Scientific American} {\bf 224/9}, 178-184 (1971).
\bibitem{Turing1936}A.~M. Turing, On computable numbers, with an application to the
{\it Entscheidungsproblem}. {\it Proc. London Math. Soc. Ser. 2}, {\bf 42}, 230 (1936), {\it ibid} {\bf 43}, 544 (1937). 
\bibitem{Chaitin97}G.~J. Chaitin, {\it The Limits of Mathematics}
(Springer, Singapore, 1997).
\bibitem{CerfAdami97}N.~J. Cerf and C. Adami, Negative entropy and
information in quantum mechanics. {\it Phys. Rev. Lett.} {\bf 79}, 5195-5197 (1997).
\bibitem{CerfAdami99}N.~J. Cerf and C. Adami, Quantum extension of
conditional probability. {\it Phys. Rev. } {\bf A 60}, 893-897 (1999).
\bibitem{AdamiCerf99}C. Adami and N.~J. Cerf, What information theory
can tell us about quantum reality. {\it Lect. Notes in Comp. Sci.} {\bf 1509}, 1637-1650 (1999).
\bibitem{AdamiCerf1997}C. Adami and N.~J. Cerf. von Neumann capacity of
noisy quantum channels. {\it Phys. Rev.} {\bf A 56}, 3470-3483 (1997).
\bibitem{Bennettetal2002}C.~H. Bennett, P.~W. Shor, J.~A. Smolin, and
A.~V. Thapliyal, Entanglement-assisted capacity of a quantum channel and the reverse
Shannon theorem. {\it IEEE Trans. Info. Theory} {\bf 48}, 2637-2655 (2002).
\bibitem{Shor94}P.~W. Shor, Algorithms for quantum computation: Discrete logarithms and factoring, in {\it Proceedings of the 35th Symposium
on Foundations of Computer Science}, edited by S. Goldwasser (IEEE Computer
Society, New York, 1994), pp. 124-134.
\bibitem{julsgaard2001}B. Julsgaard, A. Kozhekin, and
E.~S. Polzik. Experimental long-lived entanglement of two macroscopic objects. {\it
Nature} {\bf 413}, 400 (2001).
\bibitem{JarettCover1981}K. Jarett and T.~M. Cover, Asymmetries in relativistic information flow. {\it IEEE Trans. Info. Theory} {\bf 27}, 152-159 (1981). 
\bibitem{Gingrichpc}R.~M. Gingrich, unpublished (2002).
\bibitem{Aldrovandi92}R Aldrovandi and J. Gariel, On the riddle of the moving
thermometer. {\it Phys. Lett.} {\bf A 170}, 5 (1992).
\bibitem{Old}A. Einstein, \"Uber das Relativit\"atsprinzip und die aus
demselben gezogenen Folgerungen. {\it Jahrb. f. Rad. und Elekt.} {\bf 4}, 411
(1907); M. Planck, Zur Dynamik bewegter Systeme. {\it Ann. d. Phys.} {\bf 26}, 1
(1908);.
\bibitem{Ottarz6364}H. Ott, Lorentz-Transformation der W\"arme.
{\it Z. f. Physik}  {\bf 175}, 70 (1963); H. Arzeli\`es, Transformation relativiste
de la temp\'erature et de quelques autres grandeurs thermodynamiques. {\it Nuov.
Cim.} {\bf 35},792 (1964).
\bibitem{Landsberg96}P.~T. Landsberg and G.~E.~A. Matsas, Laying the ghost of the
relativistic temperature transformation. {\it Phys. Lett.} {\bf A 223}, 401 (1996).
\bibitem{Pauli21}W. Pauli, {\it Die Relativit\"atstheorie}, Encyklop\"adie der mathematischen Wissenschaften {\bf 5/2}  (Teubner, Leipzig, 1921).
\bibitem{Peebles68}P.~J.~B. Peebles and D.~T. Wilkinson. Comment on the
anisotropy of the primeval fireball. {\it Phys. Rev.} {\bf 174}, 2168 (1968).
\bibitem{Beckman01}D. Beckman, D. Gottesman, M.~A. Nielsen, and
J. Preskill, Causal and localizable quantum operations. {\it Phys. Rev.} {\bf A 64}
(2001) 052309.
\bibitem{Peres02}A. Peres, P.~F. Scudo, and D.~R. Terno, Quantum entropy
and special relativity. {\it Phys. Rev. Lett.} {\bf 88} (2002) 230402.
\bibitem{AlsingMilburn02}P.~M. Alsing and G.~J. Milburn, On Lorentz
invariance of entanglement. {\it Quant. Info. and Comp.} {\bf 2} (2002) 487-512.
\bibitem{PeresTerno03a}A. Peres and D.~R. Terno, Relativistic
Doppler effect in quantum communication. {\it J. Mod. Optics} {\bf 49} (2003)
1255-1261.
\bibitem{BDSW96}C.~H. Bennett, D.~P. DiVincenzo, J.~A. Smolin, and
W.K. Wootters, Mixed state entanglement and quantum error correction. {\it Phys.
Rev.} {\bf A 54} (1996) 3824-3851.
\bibitem{Czachor1997}M. Czachor, Einstein-Podolsky-Rosen experiment with
relativistic massive particles. {\it Phys. Rev.} {\bf A 55}, 72-77 (1997).
\bibitem{Wootters98}W.~K. Wootters, Entanglement of formation of an
arbitrary state of two qubits. {\it Phys. Rev. Lett.} {\bf 80}, 2245 (1998).
\bibitem{GingrichAdami02}R.~M. Gingrich and C. Adami, Quantum
entanglement of moving bodies. {\it Phys. Rev. Lett.} {\bf 89} (2002) 270402.
\bibitem{GingrichBergouAdami03}R.~M. Gingrich, A.~J. Bergou, and
C. Adami, Entangled light in moving frames. {\it Phys. Rev.} {\bf A 68} (2003)
042102.
\bibitem{PeresTerno03b}A. Peres and D.~R. Terno, Quantum information
and relativity theory. {\it Rev. Mod. Phys.} {\bf 75}, 93 (2004).
\bibitem{FultonRohrlich1960}T. Fulton and F. Rohrlich, Classical radiation from a uniformly accelerated charge. {\it Ann. Phys.} {\bf 9}, 499 (1960).
\bibitem{FrolovGinzburg1986}V.~P. Frolov and V.~L. Ginzburg, Excitation and radiation of an accelerated detector and anomalous Doppler effect. {\it Phys. Lett.} {\bf 116}, 423 (1986). 
\bibitem{BirrellDavies1982} 
 N.~D. Birrell and P.~C.~W. Davies, {\it Quantum Fields in Curved Space} (Cambridge University Press, Cambridge, 1982). 
\bibitem{Fulling1973}S.~A. Fulling, Non-uniqueness of canonical field quantization
in Riemannian space-time. {\it Phys. Rev.} {\bf D7}, 2850 (1973).
\bibitem{Davies1975}P.~C.~W. Davies, Scalar particle production in Schwarzschild and
Rindler metrics. {\it J. of Physics} {\bf A 8}, 609 (1975).
\bibitem{Unruh1976} W.~G. Unruh, Notes on black-hole evaporation. {\it Phys. Rev.}
{\bf D 14}, 870 (1976).
\bibitem{Alsingetal2003}P.~M. Alsing, D. McMahon, and G.~J. Milburn, Teleportation in
a non-inertial frame. LANL preprint quant-ph/0311096 (2003).
\bibitem{Hawking1975}S.~W. Hawking, Particle creation by black holes.
{\it Commun. Math. Phys.} {\bf 43} (1975) 199; Black holes and thermodynamics. {\it
Phys. Rev.} {\bf D 13} (1976) 191.
\bibitem{Preskill}J. Preskill, Do black holes destroy information?, in
  {\it Proceedings of the International Symposium on Black Holes,
    Membranes, Wormholes and Superstrings}, S. Kalara and D.V.
  Nanopoulos, eds. (World Scientific, Singapore, 1993) pp. 22-39.; T.
  Banks, Lectures on black holes and information loss. {\it Nucl. Phys. B}
  (Proc. Suppl.) {\bf 41} (1995) 21.
\bibitem{Bekenstein1973}J.~D. Bekenstein, Black holes and entropy. {\it Phys. Rev.} {\bf D 7}, 2333 (1973).
\bibitem{KokYurtsever2003}P. Kok and U. Yurtsever, Gravitational decoherence. {\it Phys. Rev.} {\bf D 68}, 085006 (2003).
\bibitem{Wald1994}R.~M. Wald, {\it Quantum Field Theory in Curved Spacetime and Black Hole Thermodynamics}. (Chicago University Press, Chicago, 1994). 
\bibitem{BekensteinMeisels1977}
J.~D. Bekenstein and A. Meisels, Einstein A and B coefficients for a black hole.
{\it Phys. Rev.} {\bf  D 13}, 2775 (1977).
\bibitem{Wald1976}
R.~M. Wald, Stimulated-emission effects in particle creation near black holes.
 {\it Phys. Rev.} {\bf D 13}, 3176 (1976),
\bibitem{AudretschMueller1992}
J. Audretsch and R. M{\" u}ller, Amplification of the black-hole Hawking radiation by stimulated emission. {\it Phys. Rev.} {\bf D 45}, 513 (1992).
\bibitem{Schiffer1993}
M.~Schiffer, Is it possible to recover information from the black-hole radiation? 
{\it Phys. Rev.} {\bf  D 48}, 1652 (1993).
\bibitem{MuellerLousto1994}
R. M{\"u}ller and C.~O. Lousto, Recovery of information from black hole radiation by considering stimulated emission. {\it Phys. Rev.} {\bf D 49}, 1922 (1994). 
\bibitem{Holevo2002} A.~S. Holevo, On entanglement-assisted classical capacity. {J. Math. Phys.} {\bf 43}, 4326 (2002).
\bibitem{FrolovPage1993}V.~P. Frolov and D.~N. Page, Proof of the generalized second law for quasistationary black holes. {\it Phys. Rev. Lett.} {\bf 71}, 3902 (1993).
\bibitem{AdamiVerSteeg2004}C. Adami and G.~L. Ver Steeg, forthcoming. 
\bibitem{Bombellietal1986}L. Bombelli, R.~K. Koul, J. Lee, and R.~D. Sorkin, Quantum source of entropy for black holes. {\it Phys. Rev.} {\bf D 34}, 373 (1986).
\bibitem{CallanWilczek1994}C. Callan anf F. Wilczek, On geometric entropy. {\it Phys. Lett.} {\bf B 333}, 55 (1994).
\bibitem{Holzhey94}C. Holzhey, F. Larsen, and F. Wilczek, Geometric
and renormalized entropy in conformal field theory. {\it Nucl. Phys.} {\bf B 424}
(1994) 443.
\bibitem{Hawking77}S.~W. Hawking, Zeta function renormalization of path
integrals in curved space time. {\it Commun. Math. Phys.} {\bf 55} (1977) 133.
\bibitem{Terno2004} D.~R. Terno, Entropy, holography and the second law. LANL preprint hep-th/0403142 (2004).
\end{thebibliography}
\end{document}